\documentclass[aps, pra, twocolumn, 10pt, floatfix, control, superscriptaddress, nofootinbib, tightenlines]{revtex4-2}

\usepackage[final]{graphicx}
\usepackage{times,bbm,amsmath,amssymb,amsthm}
\usepackage{epsfig,color}
\usepackage[dvipsnames]{xcolor}
\usepackage[colorlinks=true,citecolor=blue,linkcolor=blue]{hyperref}
\hypersetup{
    allcolors  = {blue},
}

\usepackage{
cleveref}
\usepackage{float,siunitx}
\usepackage[caption = false]{subfig}
\usepackage{verbatim}
\usepackage[greek,english]{babel}
\usepackage{thumbpdf,enumerate}
\usepackage{booktabs}
\usepackage{sidecap}
\usepackage[scaled=.8]{couriers}    
\usepackage{pstricks}
\usepackage{multirow}
\usepackage{placeins}
\usepackage{relsize}  
\usepackage{pst-grad,bm}
\usepackage{epigraph}
\usepackage{gensymb}
\usepackage{longtable}
\usepackage{booktabs}
\usepackage{gensymb}
\usepackage{mathtools}

\usepackage{soul}
\usepackage{ulem} 
\usepackage{acronym}
\usepackage{physics}
\usepackage{tikz}

\begin{document}

\author{Giovanni Rodari}
\address{Dipartimento di Fisica, Sapienza Universit\`{a} di Roma, Piazzale Aldo Moro 5, I-00185 Roma, Italy}

\author{Carlos Fernandes}
\address{International Iberian Nanotechnology Laboratory (INL) Av. Mestre José Veiga s/n, 4715-330 Braga, Portugal}

\author{Eugenio Caruccio}
\address{Dipartimento di Fisica, Sapienza Universit\`{a} di Roma, Piazzale Aldo Moro 5, I-00185 Roma, Italy}

\author{Alessia Suprano}
\address{Dipartimento di Fisica, Sapienza Universit\`{a} di Roma, Piazzale Aldo Moro 5, I-00185 Roma, Italy}

\author{Francesco Hoch}
\address{Dipartimento di Fisica, Sapienza Universit\`{a} di Roma, Piazzale Aldo Moro 5, I-00185 Roma, Italy}

\author{Taira Giordani}
\address{Dipartimento di Fisica, Sapienza Universit\`{a} di Roma, Piazzale Aldo Moro 5, I-00185 Roma, Italy}

\author{Gonzalo Carvacho}
\address{Dipartimento di Fisica, Sapienza Universit\`{a} di Roma, Piazzale Aldo Moro 5, I-00185 Roma, Italy}

\author{Riccardo Albiero}
\address{Istituto di Fotonica e Nanotecnologie, Consiglio Nazionale delle Ricerche (IFN-CNR), Piazza Leonardo da Vinci 32, I-20133 Milano, Italy}

\author{Niki Di Giano}
\address{Istituto di Fotonica e Nanotecnologie, Consiglio Nazionale delle Ricerche (IFN-CNR), Piazza Leonardo da Vinci 32, I-20133 Milano, Italy}
\address{Dipartimento di Fisica, Politecnico di Milano, Piazza Leonardo da Vinci 32, 20133 Milano, Italy}

\author{Giacomo Corrielli}
\address{Istituto di Fotonica e Nanotecnologie, Consiglio Nazionale delle Ricerche (IFN-CNR), Piazza Leonardo da Vinci 32, I-20133 Milano, Italy}

\author{Francesco Ceccarelli}
\address{Istituto di Fotonica e Nanotecnologie, Consiglio Nazionale delle Ricerche (IFN-CNR), Piazza Leonardo da Vinci 32, I-20133 Milano, Italy}

\author{Roberto Osellame}
\address{Istituto di Fotonica e Nanotecnologie, Consiglio Nazionale delle Ricerche (IFN-CNR), Piazza Leonardo da Vinci 32, I-20133 Milano, Italy}

\author{Daniel J. Brod}
\address{Instituto de F\'{i}sica, Universidade Federal Fluminense, Niter\'{o}i -- RJ, Brazil}

\author{Leonardo Novo}
\address{International Iberian Nanotechnology Laboratory (INL) Av. Mestre José Veiga s/n, 4715-330 Braga, Portugal}

\author{Nicol\`o Spagnolo}
\email{nicolo.spagnolo@uniroma1.it}
\address{Dipartimento di Fisica, Sapienza Universit\`{a} di Roma, Piazzale Aldo Moro 5, I-00185 Roma, Italy}

\author{Ernesto F. Galv\~{a}o}
\email{ernesto.galvao@inl.int}
\address{International Iberian Nanotechnology Laboratory (INL) Av. Mestre José Veiga s/n, 4715-330 Braga, Portugal}
\address{Instituto de F\'{i}sica, Universidade Federal Fluminense, Niter\'{o}i -- RJ, Brazil}

\author{Fabio Sciarrino}
\address{Dipartimento di Fisica, Sapienza Universit\`{a} di Roma, Piazzale Aldo Moro 5, I-00185 Roma, Italy}

\title{Experimental observation of counter-intuitive features of photonic bunching}

\begin{abstract}
Bosonic bunching is a term used to describe the well-known tendency of bosons to bunch together, and which differentiates their behaviour from that of fermions or classical particles. However, in some situations perfectly indistinguishable bosons may counter-intuitively bunch less than classical, distinguishable particles. Here we report two such counter-intuitive multiphoton bunching effects observed with three photons in a three-mode balanced photonic Fourier interferometer. In this setting, we show indistinguishable photons actually minimize the probability of bunching. We also show that any non-trivial value of the three-photon collective photonic phase leads to a decreased probability of all photons ending up in the same mode, even as we increase pairwise indistinguishability. Our experiments feature engineering of partial indistinguishability scenarios using both the time and the polarization photonic degrees of freedom, and a polarization-transparent 8-mode tunable interferometer with a quantum-dot source of single photons. Besides the foundational understanding, the observation of these counter-intuitive phenomena open news perspective in devising more efficient ways of routing photons for advantage in metrology and quantum computation.
\end{abstract}

\maketitle

\section {Introduction}
A fundamental consequence of bosonic statistics is the tendency for bosons to occupy the same state, leading to physical phenomena like Bose-Einstein condensation \cite{bloch2022BEC} or the Hong-Ou-Mandel effect \cite{hong1987measurement,bouchard2020two}. In the latter, two identical photons entering different input arms of a balanced beam splitter always exit via the same output arm, due to constructive interference between the two indistinguishable paths that make the two photons coalesce. It is well known that any distinguishability between the input photons (e.g. in the frequency or polarization degrees of freedom) provides which-path information, consequently degrading quantum interference effects and resulting in a lower probability for bosons to bunch.

In multiphoton interference experiments, several ways to quantify boson bunching have been proposed \cite{spagnolo2013general, carolan2014experimental,shchesnovich2016bunching} and the precise connection between bosonic bunching and the distinguishability of the photons is still poorly understood. While it is generally predicted and experimentally observed that any amount of distinguishability degrades boson bunching effects \cite{spagnolo2013general, carolan2014experimental,shchesnovich2016bunching, niu2009observation, fourphotonHOM}, recent counterexamples have been found. In Refs. \cite{Seron23,pioge2023enhanced} it was theoretically shown that multimode bunching probabilities, i.e. the probability for all photons to coalesce in a subset of output modes, are not always maximized by perfectly indistinguishable photons. Such anomalous behaviour of boson bunching was predicted to occur only in interference experiments with 7 or more photons, leaving open the question of whether other such counter-intuitive features of boson bunching can be observed in smaller experiments. 

In this work we study the role of photon indistinguishability in two natural measures of boson bunching, thus observing their counter-intuitive behaviour in three-photon interference experiments. The first measure of boson bunching we consider, which we simply refer to as bunching probability, is the probability of observing outcomes with two or more photons in the same output mode. This is also referred to as the probability of collisions in Boson Sampling experiments and, when averaged over random interferometers, it is known to be larger for indistinguishable bosons than for distinguishable particles \cite{aaronson2011computational, bosonbirthdayparadox}. We demonstrate that this average behaviour is not always valid. Focusing on the balanced tritter interferometer, we show not only that distinguishability can increase the bunching probability but that this quantity is, in fact, \textit{minimized} if the photons are fully indistinguishable. 

The second measure of boson bunching we consider is the full bunching probability, defined as the probability that all photons exit the interferometer in a given output port. It is known that indistinguishable photons have an exponential enhancement of full bunching in comparison with distinguishable particles \cite{tichy2015sampling, spagnolo2013general}. While other outcome probabilities are known to have a non-monotonic behavior with respect to the pairwise indistinguishability of the photons \cite{tichy2011four, ra2013nonmonotonic}, full bunching probabilities are widely assumed to decrease monotonically with distinguishability, a claim that is supported by experimental observations \cite{fourphotonHOM,niu2009observation}.  We show this in fact is only true in restricted set-ups, where inner products between the functions describing the internal degrees of freedom of each input photons (frequency, polarization, etc.) are given by real, non-negative values. In full generality, the relational properties of the input quantum states are described by complex Gram matrices \cite{oszmaniec2024measuring}. The complex phases needed to describe such matrices can be related to many different non-classical effects \cite{Wagner2023simple, wagner2023quantum, oszmaniec2024measuring}. In the context of multiphoton interference, they are referred to as collective photonic phases \cite{shchesnovich2018collective} and only recently they have been measured in three- and four-photon  experiments \cite{menssen2017distinguishability, jones2020multiparticle, Pont22}. In our work, we unveil the role of collective photonic phases in the phenomenon of boson bunching, showing how they non-trivially affect the interference between paths that lead to all photons exiting in the same output mode. As a striking consequence of this, we experimentally demonstrate situations where full bunching probabilities can increase even though the pairwise distinguishability between the photons is decreased. In particular, we experimentally observe such counter-intuitive features in an advanced photonic platform based on a quantum-dot source and an integrated fully-reconfigurable universal and polarization independent interferometer fabricated via the femtosecond laser micromachining technology. This is performed by exploiting the peculiar property of the employed interferometer, which enables the use of multiple internal degrees of freedom to prepare classes of states suitable for the analysis of the role of collective phases in multiphoton bunching properties.

\section{Multiphoton indistinguishability and bunching}

The set-up we consider consists of an $n$-mode general linear-optical interferometer, with one single photon per input mode. The interferometer is described by a unitary $n \times n$ matrix $U$, which maps input creation operators to output creation operators. The indistinguishability of the photons is completely described by a Gram matrix of inner products: $G_{i,j}=\left\langle \psi_i |\psi_j \right\rangle$, where $\ket{\psi_i}$ is the spectral function describing all the internal degrees of freedom of the photon entering input port $i$ \cite{tichy2015sampling}. This Gram matrix can be written only in terms of physically relevant, unitary invariant properties of the set of input spectral functions, as described in \cite{oszmaniec2024measuring}. As the case with mixed states describing the internal degrees of freedom is more intricate \cite{Jones2023disting}, here we will focus on the case of pure spectral functions.

In what follows we describe in more detail the case of $n=3$ photons, as it is the simplest case where multiphoton effects play a role. As discussed in \cite{oszmaniec2024measuring}, the indistinguishability scenario for a 3-photon experiment is completely characterized by a Gram matrix:
\begin{align}
G =& 
\begin{pmatrix}
1 & \left|\left\langle \psi_1 |\psi_2 \right \rangle\right| & \left|\left\langle \psi_1 |\psi_3 \right \rangle\right| \notag \\
\left|\left\langle \psi_1 |\psi_2 \right \rangle\right| & 1 & \left\langle \psi_2 |\psi_3 \right \rangle \\
\left|\left\langle \psi_1 |\psi_3 \right \rangle\right| & \left\langle \psi_2 |\psi_3 \right \rangle^*& 1
\end{pmatrix} \\
=&
\begin{pmatrix}
1 & \sqrt{\Delta_{12}} & \sqrt{\Delta_{13}} \\
\sqrt{\Delta_{12}} & 1 & \sqrt{\Delta_{23}}e^{i\varphi} \\
\sqrt{\Delta_{13}} & \sqrt{\Delta_{23}}e^{-i\varphi}& 1
\end{pmatrix},
\label{eqn:Gram}
\end{align}
parameterized in terms of the two-photon overlaps $\Delta_{ij}=\left|\left\langle \psi_i |\psi_j \right \rangle \right|^2$ and the phase $\varphi$ of the single nontrivial 3-photon Bargmann invariant 
\begin{equation}
    \Delta_{123}= \braket{\psi_1}{\psi_2}\braket{\psi_2}{\psi_3}\braket{\psi_3}{\psi_1} = \sqrt{\Delta_{12}\Delta_{13}\Delta_{23}}e^{i\varphi}.
    \label{eqn:bargmann}
\end{equation}
The phase of $\Delta_{123}$ is known as a collective photonic phase \cite{menssen2017distinguishability}, and is not accessible in experiments with fewer than 3 photons. We note that the phase $\varphi$ cannot be chosen arbitrarily, it was recently shown that the value of $|\Delta_{123}|$ imposes constraints on the allowed values for $\varphi$, stemming from the fact that a Gram matrix is positive-semidefinite \cite{2024unitaryinvariant} (see also Supplementary Note 1).  

\begin{figure}[ht!]
    \centering
    \includegraphics[width=0.425\textwidth]{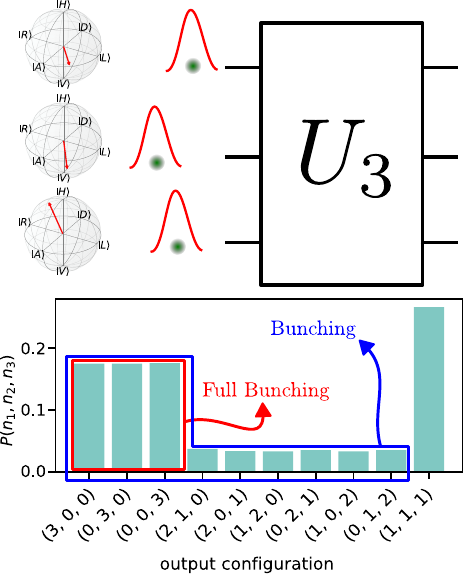}
    \caption{\textbf{Characterization of the quantum interference of 3 photons evolving in a 3-mode interferometer.}  Three photons are injected in a linear optical network $U_3$ implementing the 3-mode Fourier transform, and we sample from the distribution $p(n_1,n_2,n_3)$ of photon numbers at the output modes. From this distribution we extract the bunching probability $p_{\text{B}}$ and the full bunching probability $p_{\text{FB}}$, as shown in the lower panel. The polarization and time delays of the photons are manipulated to prepare indistinguishability scenarios described by a 3x3 Gram Matrix $G$.}
    \label{fig:conceptual}
\end{figure}

Here we will investigate some counter-intuitive behaviors of two different figures of merit describing multiphoton bunching:
\begin{itemize}
    \item Probability of bunching $p_{\text{B}}$ (or collisions), defined as the probability that at least one output mode has two or more photons.
    \item Probability of full bunching $p_{\text{FB}}^{(i)}$ in the $i$th  output mode -- the probability that all photons come out of the interferometer in output mode $i$. 
\end{itemize}
We will focus on analyzing these probabilities for the balanced tritter, i.e. a Fourier interferometer of three modes,  with 1 photon per input mode (see Fig.~\ref{fig:conceptual}). 
 For a given indistinguishability scenario described by a Gram matrix of the form of Eq.~\eqref{eqn:Gram}, the bunching and full bunching probabilities depend only on the 3-photon Bargmann invariant and the average two-photon overlap 
\begin{equation}
    \bar{\Delta}=\frac{\Delta_{12}+\Delta_{23}+\Delta_{13}}{3},
    \label{eq:av_indist}
\end{equation}
and are given by:
\begin{eqnarray}
    p_{\text{B}} &=& 1 + \frac{1}{9}(3 \bar{\Delta} - 4 \cos{\varphi} |\Delta_{123}|-2 )\label{eq:pb}\\
    p_{\text{FB}}^{(i)} &=& \frac{1}{27}(1+ 3\bar{\Delta}+ 2 \cos{\varphi}|\Delta_{123}|),\label{eq:pfb}
\end{eqnarray}
for any output mode  $i\in \{1,2,3\}$ (see Ref. \cite{menssen2017distinguishability}).

\textit{\textbf{Counter-intuitive behaviour of $p_\text{B}$}}. There are choices of interferometers for which partial indistinguishability scenarios have \textit{higher} bunching probabilities $p_{\text{B}}$ than the fully indistinguishable case. The most extremal case we found for 3-mode interferometers occurs for the balanced tritter which is implemented experimentally in this work. In this case, $p_{\text{B}}$ is actually \textit{minimized} by perfectly indistinguishable photons. Hence, any increase in distinguishability (without necessarily requiring a non-trivial collective photonic phase) will increase the bunching probability.  This is in sharp contrast with the expected behavior for Haar random matrices where, for example, distinguishable photons tend to bunch less than indistinguishable ones~\cite{bosonbirthdayparadox}. While counterexamples to this behavior can be found by generating random interferometers of small dimensions, this counter-intuitive behavior becomes rarer as the dimension grows (see Supplementary Note 2). It is also interesting to point out that for Fourier interferometers of even dimension with indistinguishable photons at the input, all allowed outcomes lead to bunching due to suppression laws \cite{lim2005generalized, tichy2010zero}, meaning that full indistinguishability maximizes bunching in this case. Further examples of counter-intuitive bunching phenomena with a higher number of modes can be found for Fourier interferometers of certain odd dimensions, or real Hadamard interferometers with entries $\pm 1/\sqrt{n}$ which extremize the permanent \cite{wanless2005permanents} (see Supplementary Note 2). 

Going back to the three-mode Fourier interferometer, which is implemented experimentally in this work, it is natural to ask what kind of distinguishability scenario maximizes bunching. Using Eq.~\eqref{eq:pb}, it can be seen that maximum probability of bunching is obtained for the largest possible negative value of the three-photon Bargmann invariant, a scenario characterized by equal overlaps $\Delta_{ij}=1/4$ and a collective photonic phase of $\pi$, which we explore experimentally in Section \ref{sec:experimental}. We refer to Supplementary Note 3 for a detailed derivation of the maximum and minimum bunching probabilities for the three-mode Fourier interferometer.  

\textit{\textbf{Counter-intuitive behaviour of $p_{\text{FB}}$}}
It is known that full bunching probabilities $p_{\text{FB}}^{(i)}$ are always maximized for fully indistinguishable photons \cite{tichy2015sampling}, representing an intuitive behaviour. However, $p_{\text{FB}}$ fails to behave in an intuitive way in certain partial indistinguishability scenarios. This follows from a simple result due to Tichy \cite{tichy2015sampling}, relating full bunching probabilities in two different distinguishability scenarios described by Gram matrices $G_1,G_2$:
\begin{equation}
\frac{p_{\text{FB}}^{(i)}(G_1)}{p_{\text{FB}}^{(i)}(G_2)}= \frac{\text{Per}(G_1)}{\text{Per}(G_2)}, \label{eq:fb}
\end{equation}
where $\text{Per}(\cdot)$ denotes the matrix permanent function. The calculation of such permanents involves the product of the Gram matrix elements, thus leading in the 3-photon case to an expression for $p_{\text{FB}}^{(i)}$ depending on the third-order Bargmann invariation as in Eq. \eqref{eq:pfb}. Additionally, Eq. (\ref{eq:fb}) is a more general version of the full-bunching law discussed in \cite{spagnolo2013general}, that applies also to partial indistinguishability scenarios. The result above is independent of which interferometer is being used, and of the chosen output mode.

If we restrict ourselves to Gram matrices with non-negative entries the intuitive behavior of full bunching is recovered: any decrease of two-photon overlaps $\Delta_{ij}= |\braket{\psi_i}{\psi_j}|^2$ results in diminished full bunching probabilities. However, the explicit dependency of $\text{Per}(G)$ on the complex phases appearing in the Gram matrix, as can be seen from Eqs.~(\ref{eq:pfb}-\ref{eq:fb}), clearly shows that pairwise distinguishability is not sufficient to understand full bunching phenomena. In fact, for three photons, if we fix the values of the overlaps $\Delta_{ij}$, a nontrivial collective phase $\varphi$ \emph{decreases} full bunching probabilities when compared to the same scenario with $\varphi=0$. As a consequence of this,  for any $\varphi>0$, it is always possible to find scenarios where pairwise indistinguishability decreases but at the same time full bunching probabilities increase. More precisely, it follows from Eqs.~(\ref{eq:pfb}-\ref{eq:fb}) that a sufficiently small decrease of pairwise indistinguishability, obtained by lowering the values of $\bar{\Delta}$ and $|\Delta_{123}|$, can be compensated in the full bunching probability by tuning the collective phase $\varphi$ towards 0. This effect is of course more prominent when we are allowed large variations of $\varphi$. For this reason, in the experimental sections, we focus on distinguishability scenarios targeting the preparation of states with $\Delta_{ij} = 1/4$ which allows for arbitrary changes of the phase between $[-\pi, \pi]$ \cite{2024unitaryinvariant}, to verify the theoretical predictions. 

Interestingly, the precise condition $\varphi \neq 0$ which we associate to counter-intuitive features of full bunching is also associated with nonclassicality in other settings. As examples, the appearance of a non-trivial phase in a Bargmann invariant describing each scenario is required for anomalous weak values \cite{Wagner2023simple}, for nonclassical Kirkwood-Dirac quasiprobabilities \cite{wagner2023quantum}, and serves as a basis-independent witness of coherence \cite{oszmaniec2024measuring}. 

\section{Experimental Verification}
\label{sec:experimental}

\begin{figure*}[ht!]
    \centering
    \includegraphics[width=0.99\linewidth]{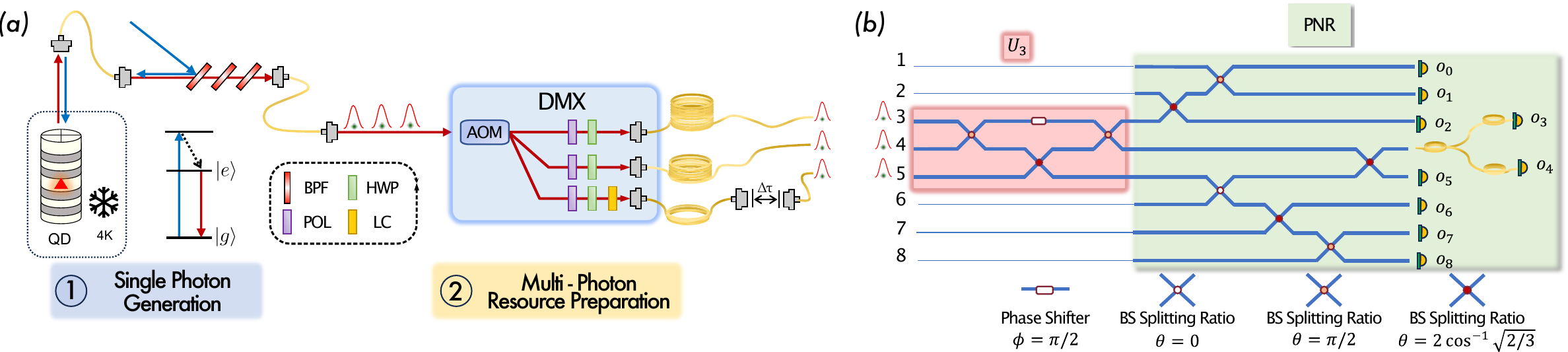}
    \caption{\textbf{Experimental platform for the verification of counter-intuitive bunching properties:} in this experimental verification, we employ the QOLOSSUS machine, consisting of several interconnected stages. a) A Quantum Dot (QD) based single-photon source is excited in a non-resonant, longitudinal acoustic phonon assisted scheme, operating at cryogenic temperatures, and generating a stream of single photons at regular time intervals with a repetition rate of $79$ MHz. Then, via a time-to-spatial demultiplexer (DMX) employing an acousto-optical modulator, the train of single photons is divided into three spatially separated modes, which are then temporally synchronized via properly tuned in-fiber delay loops. Inside the bulk geometry of the DMX setup, in order to independently control the spectral function $\ket{\psi_i}$ of each of the photons in our resource state, we add a polarizer and a half-wave plate for each channel (purple and green rectangles). Moreover, a liquid crystal (yellow rectangle) and a time-delay line are added to one of the channels to obtain full control of both time and polarization degrees of freedom. 
    b) The prepared states are then injected in an eight mode, fully reconfigurable photonic integrated circuit. The circuit is programmed in two separate blocks. The red module is programmed as a three-mode balanced Fourier interferometer described by unitary $\Tilde{U}_3$ where interference occurs. Then each of its three output modes are manipulated separately in the green block where, with an additional in-fiber BS, we implement a Pseudo-Number Resolving (PNR) measurement in order to reconstruct the full photon number output distribution $p(n_1,n_2,n_3)$. \textit{Legend - } QD - Quantum Dot, BPF - Band Pass Filter, POL - Linear Polarizer, HWP - Half Wave Plate, LC - Liquid Crystal Retarder, AOM - Acousto-Optical Modulator.}
    \label{fig:experimental_setup}
\end{figure*}

We now provide an experimental verification of counter-intuitive effects due to multiphoton quantum interference, which occur when more than two photons propagate within a multimode optical interferometer.
The experiment has been carried out employing the QOLOSSUS photonic machine introduced in \cite{rodari2024semi}, whose setup is shown in Fig.~\ref{fig:experimental_setup} and described in more detail in the Methods section and in Supplementary Note 4. Briefly, the experimental setup can be divided into three sequential stages, related to single-photon generation via a Quantum Dot (QD) based source \cite{heindel2023quantum, senellart2017high,gazzano2013bright,somaschi2016near,ollivier2020reproducibility, thomas2021bright}; multi-photon state preparation with a bulk time-to-spatial demultiplexing setup (DMX) \cite{anton2019interfacing, Pont22, pont2022high, rodari2024semi}; and state evolution with pseudo-photon-number resolved detection implemented via a reconfigurable integrated photonic processor (IPP) \cite{Clements2016, flamini2015thermally, pentangelo2024high}.
Indeed, in order to engineer 3-photon states described by arbitrary Gram matrices, one can independently prepare each quantum state $\ket{\psi_i}$ using either polarization or time-delay. By introducing a set of half-wave plates, and an additional liquid crystal plus a time-delay on one of the photons (see Fig. \ref{fig:experimental_setup}a), one can prepare a set of three-photon states with the general form:
\begin{equation}
    \begin{split}
        \ket{\psi_1} & = \ket{0}_t \otimes (\cos(\alpha)\ket{0}_p + \sin(\alpha)\ket{1}_p) \\
        \ket{\psi_2} & = \ket{0}_t \otimes (\cos(\beta)\ket{0}_p + \sin(\beta)\ket{1}_p) \\
        \ket{\psi_3} & = (x\ket{0}_t+\sqrt{1-x^2}\ket{1}_t) \otimes \\
        & \otimes (\cos(\gamma)\ket{0}_p + \sin(\gamma) e^{i\phi} \ket{1}_p) \\
    \end{split}
    \label{eqn:stateprep}
\end{equation}
where $\{\ket{0}_p, \ket{1}_p \}$ and $\{\ket{0}_t, \ket{1}_t \}$ are orthogonal states in the polarization and time-delay basis, ($\alpha$, $\beta$, $\gamma$, $\phi$) are related to the orientation of the polarization states in the Bloch sphere, and $x$ is associated to the relative time-delay. After preparing the input state, the photons are injected in a three-mode balanced Fourier interferometer implemented by suitably programming an 8-mode fully reconfigurable universal integrated chip \cite{flamini2015thermally, pentangelo2024high, Clements2016}. The effective three-mode unitary implemented on our device, reported in the Methods, has a trace fidelity of $\mathcal{F} = \vert \Tr\{U_3^\dag \Tilde{U}_3\} \vert/3 = 0.99922(4)$ with respect to the ideal transformation. By employing a pseudo-number resolved detection setup (see Fig. \ref{fig:experimental_setup}b), we are able to fully reconstruct the output probability distribution $p(n_1,n_2,n_3)$ and thereby recover both the total \textit{full-bunching} probability $p_{\text{FB}}$, associated with a given three-photon Gram-matrix $G$:
\begin{equation}
    p_{\text{FB}}(G) = p(3,0,0;G) + p(0,3,0;G) + p(0,0,3;G)
\end{equation}
and the \textit{bunching} probability:
\begin{equation}
    p_{\text{B}}(G) = 1 - p(1,1,1;G)
\end{equation}
(see Fig. \ref{fig:conceptual}), where with $G$ we indicated explicitly the dependence on the Gram matrix.

\begin{figure*}[ht!]
    \centering
    \includegraphics[width=0.99\linewidth]{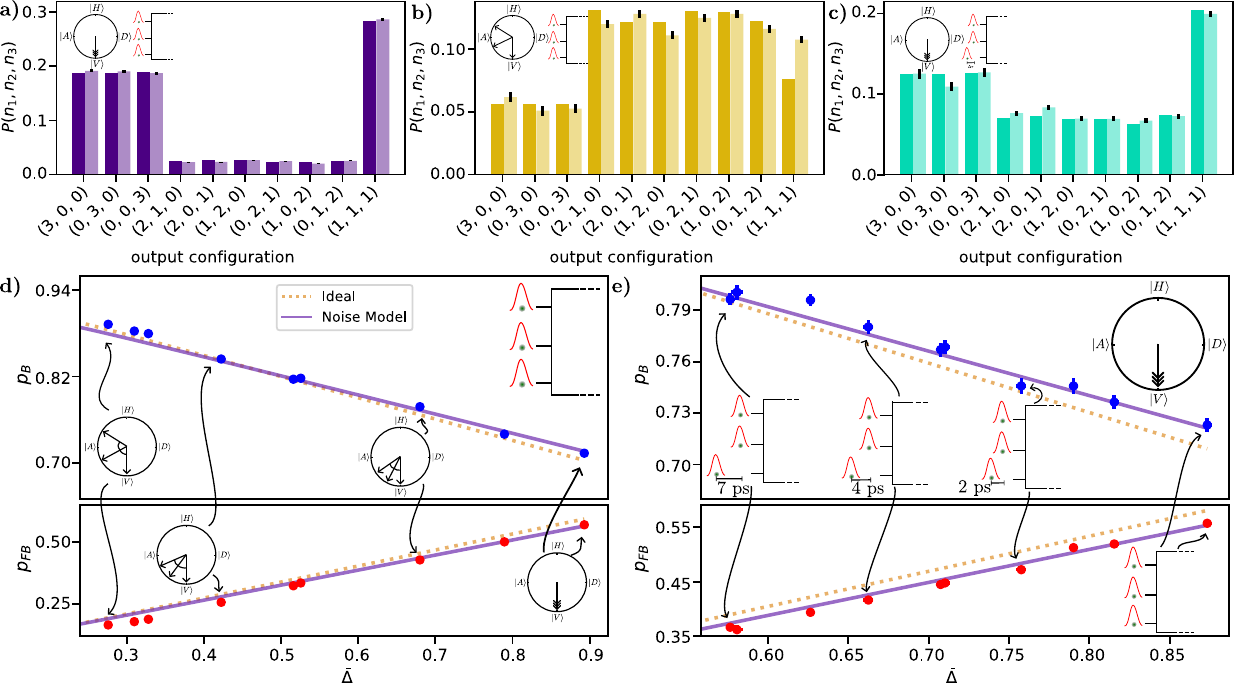}
    \caption{\textbf{Measured and expected probabilities of bunching $p_{\text{B}}$ and full bunching $p_{\text{FB}}$, with control of photon polarization or time delay.} The measured and simulated probability distributions $p(n_1,n_2,n_3)$, over the output Fock states, are shown: \textbf{a)} in the nearly indistinguishable case; in the case in which \textbf{b)} only the photon polarizations are manipulated by half-wave plates in order to obtain three states in the $x-z$ plane of the Bloch sphere; and \textbf{c)} only an adjustable time delay is applied to one channel. The probabilities of bunching $p_{\text{B}}(G) = 1 - p(1,1,1;G)$ and full bunching $p_{\text{FB}}=p(3,0,0;G) + p(0,3,0;G) + p(0,0,3;G)$ are obtained from the measured and simulated distributions, as a function of the average overlap $\bar{\Delta}$ which varies according to the \textbf{d)} polarization and \textbf{e)} time delay configuration. Here the (dashed) lines represent a (ideal) model of bunching and full-bunching expected at the output, with either a polarization-based or time-based Gram matrix preparation. As highlighted in the Supplementary Material, we can account for sources of experimental noise such as imperfections in the three-mode unitary implementation and a non-unit single photon purity of the source, with good agreement between experiment and the numerical prediction.}
    \label{fig:bunching and full}
\end{figure*}

In what follows, we report results associated with different families of Gram matrices engineered by suitably preparing input states as in \eqref{eqn:stateprep}. We note that the overall quantum state prepared in our setup will depend on all internal degrees of freedom of the photon states. In other words, one can write the internal state of each photon $i$ as $\ket{\psi_i} = \ket{\chi_i}_o \otimes \ket{\mu_i}_t \otimes \ket{\nu_i}_p$. Here, $\ket{\chi}_o$ is associated to the properties of the quantum dot emission itself and related to the maximum pairwise overlap achievable between photon pairs when they are aligned both in time and polarization, i.e. $\Delta^{\text{max}}_{ij} = \vert  \bra{\chi_i}\ket{\chi_j}_o \vert^2$. Conversely, $\ket{\mu_i}_t$ and $\ket{\nu_i}_p$ describe the internal states in the time and polarization degrees of freedom respectively, which are tuned in the experiment. Hereafter, the pairwise overlaps $\tilde{\Delta}_{ij}$ have been estimated by measuring the visibilities $V^{\mathrm{HOM}}_{ij}$ of independent Hong-Ou-Mandel experiments carried out by suitably reconfiguring the IPP, and correcting for the bias effect introduced by the presence of a multi-photon component due to the non-zero second-order correlation function $g^{(2)}(0) \neq 0$. This effect is taken into account by using the approach of \cite{ollivier2021hong}, which leads to the following relation between overlap, visibility and second-order correlation function $\tilde{\Delta}_{ij} = [V^{\mathrm{HOM}}_{ij} + g^{(2)}(0)]/[1 - g^{(2)}(0)]$.

\textbf{\textit{Experimental observation of counter-intuitive behavior of bunching}}. The adoption of a polarization-insensitive technology for our IPP \cite{pentangelo2024high} allows us to probe multi-photon interference effects generated by manipulation of the polarization degree of freedom in the state preparation stage. We generated triads of polarization states lying on a great circle of the Bloch sphere, e.g. the plane spanned by the eigenvectors of $\sigma_x$ and $\sigma_z$. In particular, using only the half-wave plates we can prepare states with $x=1$, $\gamma = 0$ and $\beta = \alpha + \delta$ in Eq. \eqref{eqn:stateprep}. Moreover, we choose $\alpha = \delta$ leading to a real Bargmann invariant quantified as $\Delta_{123} \propto \cos(\alpha)\cos(\alpha)\cos(2\alpha)$, since with such a choice of states it can be shown that the triad phase associated with $\Delta_{123}$ is either $0$ or $\pi$. The comparison between the measured and simulated probability distributions associated with two different Gram matrices with triad phase $0$ or $\pi$ are reported in Fig. \ref{fig:bunching and full}a-b. In particular, they correspond to the nearly fully indistinguishable case, in which $\alpha=0$ and $\Delta_{ij} \approx \Delta^{\text{max}}_{ij}$ (Fig. \ref{fig:bunching and full}a), and to the case in which the pairwise overlaps are balanced, $\alpha=\pi/6$ and $\Delta_{ij} \approx \Delta^{\text{max}}_{ij}/4$ (Fig. \ref{fig:bunching and full}b). In addition to these extremal conditions, we engineered ten more Gram Matrices by varying $\alpha$ in the interval $[0,\pi/6]$. Therefore, in Fig. \ref{fig:bunching and full}d, we plot both the full bunching probability $p_{\text{FB}}(G)$ and the bunching probability $p_{\text{B}}(G)$ as a function of the average overlap $\bar{\Delta}$ associated to each experimental point and we compare the trends with a theoretical prediction, which includes the main sources of imperfections according to a detailed model \cite{Pont22,rodari2024semi} (see Supplemental Note 5 for further details). These measurements clearly show a counter-intuitive behavior of the bunching probability, since it \textit{decreases} as we increase the average two-photon overlap. 

\begin{figure*}[t!]
    \centering
    \includegraphics[width=0.99\textwidth]{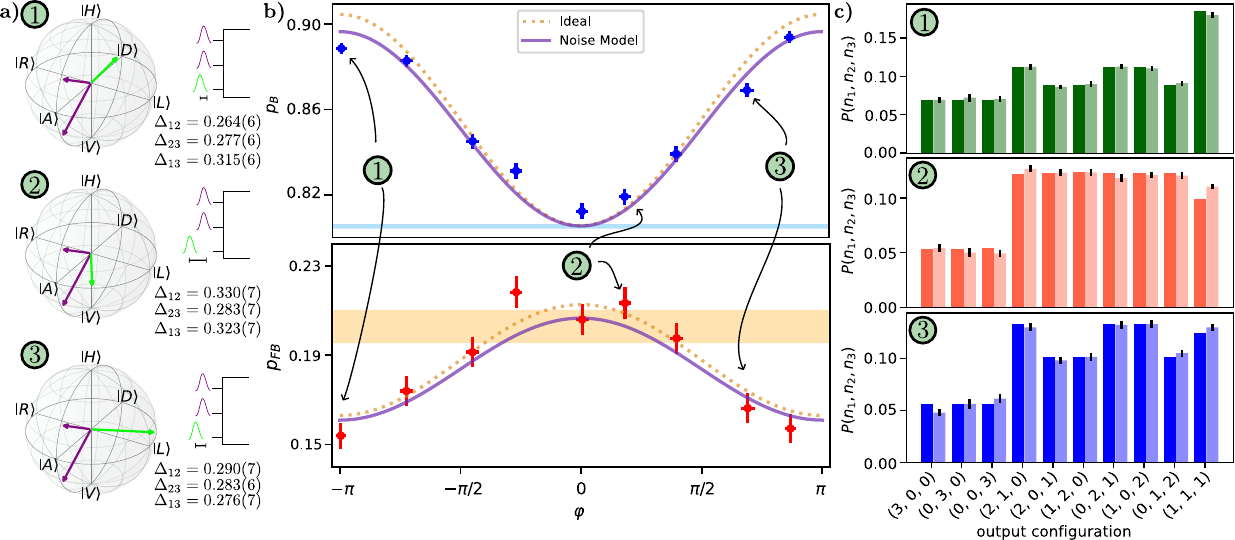}
    \caption{\textbf{Role of the triad phase on the bunching $p_{\text{B}}$ and full bunching $p_{\text{FB}}$ probabilities.} \textbf{a)} Bloch sphere visualization of three polarization configurations obtained with half-wave plates and a liquid crystal, in order to tune the relative phase between the basis states and obtain complex-valued Gram matrices; via an additional time-delay, the pairwise photon overlaps $\Delta_{ij}$ are kept fixed in such a way that $\sqrt{\Delta_{ij}} \approx 1/2$. The measured overlaps are reported. \textbf{b)} Probability of bunching $p_{\text{B}}$ and full bunching $p_{\text{FB}}$ as a function of the triad phase $\varphi$. In order to show non-trivial modulation of the (full) bunching probabilities due to a non-null triad phase, the blue and the orange regions represent the values attained by $p_{\text{B}}$ and $p_{\text{FB}}$, respectively, from numerical simulations which account for experimental imperfections and consider the experimentally measured overlaps $\Delta_{ij}$ while assuming $\varphi=0$. Again the (dashed) line represents the results of a numerical simulation of the experimentally tested Gram matrices showing a good agreement with the experimentally obtained distributions. \textbf{c)} Measured and simulated probability distributions, respectively in full and opaque colors, over the possible output Fock states $p(n_1,n_2,n_3)$ for experimental Gram matrices characterized by $\sqrt{\Delta_{ij}} \approx \frac{1}{2}$ and an increasing triad phase $\varphi$, whose values are evaluated to be $\varphi=\{-3.127,0.578,2.178\}$.}
    \label{fig:timepol}
\end{figure*}

A similar behaviour can be obtained by only manipulating the temporal degree of freedom. In particular, by preparing each of the three streams of single photons with aligned $\ket{V}$ polarization and delaying one of them with respect to the other two, we can obtain the family of states in Eq. \eqref{eqn:stateprep} corresponding to $\alpha = \beta = \gamma = 0$, while $x$ is tuned by introduction of a fine temporal delay. As argued in \cite{menssen2017distinguishability}, the manipulation of the time degree of freedom does not affect the triad phase which can only assume a null value with the use of a QD based source (see Supplementary Note 4 for more details). Varying the parameter $x$, we engineered $10$ different Gram matrices and the results are reported in Fig. \ref{fig:bunching and full}c-e. In particular, in Fig. \ref{fig:bunching and full}c we compared the experimental and simulated probability distribution for the experimental data with the lowest average indistinguishability $\bar{\Delta}= 0.576(3)$. In Fig. \ref{fig:bunching and full}e we report both the bunching and full bunching probability as a function of $\bar{\Delta}$: again, one can see that the bunching probability counter-intuitively \textit{decreases} as the average two-photon overlap increases.

\textbf{\textit{Triad phase dependence of probabilities of bunching and full bunching}.} As a next step, by acting simultaneously on both the time and polarization photonic degrees of freedom, we can highlight the effects due to a complex-valued Gram matrix. In particular, we focused on engineering a set of states (see Fig. \ref{fig:timepol}a) with $\alpha = \pi/4 + \delta$, $\beta = \pi/4 - \delta$ and $\cos(\gamma) = \sin(\gamma) = 2^{-1/2}$ and a variable phase $\phi$ as in Eq. \eqref{eqn:stateprep}. This choice of parameters, for $\delta \approx \frac{\pi}{6}$, allows us to generate the family of Gram matrices characterized by  $\sqrt{\Delta_{ij}} \approx \frac{1}{2}$ and a variable triad phase $\varphi$. This can be done by properly tuning the free parameters $x$ and $\phi$, where $x$ affects only the overlaps while $\phi$ affects also the Gram matrix phase $\varphi$. We note that the latter parameter is in a one-to-one and monotonic relationship with the polarization state's phase $\phi$, and $\phi = \varphi$ holds only if $\phi = 0, \pm \pi$.
In particular, we engineered $9$ different state triplets, with the corresponding bunching and full bunching probabilities reported in Fig. \ref{fig:timepol}b as a function of the measured triad phase. In Fig. \ref{fig:timepol}c we show the comparison between the theoretical and experimental probability distribution associated with three different conditions characterized by $\sqrt{\Delta_{ij}} \approx \frac{1}{2}$ and the triad phases $\varphi=\{-3.127,0.578,2.178\}$.
As highlighted in Fig.~\ref{fig:timepol}b, by varying the triad phase $\varphi$ carried by the Gram matrix a counter-intuitive modulation of the (full) bunching probabilities is obtained, which cannot be directly related to the knowledge of only the overlaps $\Delta_{ij}$. Indeed, the experimentally obtained (full) bunching probability deviate consistently from the shaded regions, representing the expected bunching probabilities with a real-valued Gram matrix, while also being consistent with numerical simulations carried out considering a suitable noise model of our experimental implementation.

\textbf{\textit{Experimental observation of counter-intuitive behavior of full bunching}}.  In the previous sections we have considered Gram matrices preparations describing three-photon states, as in Eq. \eqref{eqn:Gram}, manipulated both in the polarization and time degrees of freedom. This has allowed us to directly verify a counter-intuitive behaviour of the bunching probability $p_b$, i.e. the probability that at least two out of three photons exit in the same output of a balanced Fourier interferometer, showing non-trivial trends as a function of the average overlap $\bar{\Delta}$ or the triad phase $\varphi$ of the 3-photon Bargmann invariant as in Eq. \eqref{eqn:bargmann}.
We now show that, by engineering appropriately both the overlaps and the phase of a 3-photon Gram matrix, it is possible to observe a counter-intuitive behaviour arise even with respect to the {\em full-bunching} probability $p_{\text{FB}}$. More specifically, we are able to find suitable preparations $A$ and $B$ such that $p_{\text{FB}}^{(A)} \geq p_{\text{FB}}^{(B)}$ even though $\bar{\Delta}^{(A)} \leq \bar{\Delta}^{(B)}$ and $|\Delta_{123}^{(A)}| \leq |\Delta_{123}^{(B)}|$. We start with the same state prepared in the previous section, i.e. with $\alpha = \pi/4 + \delta$, $\beta = \pi/4 - \delta$ and $\cos(\gamma) = \sin(\gamma) = 2^{-1/2}$.
\begin{figure}[ht!]
    \centering
    \includegraphics[width=\columnwidth]{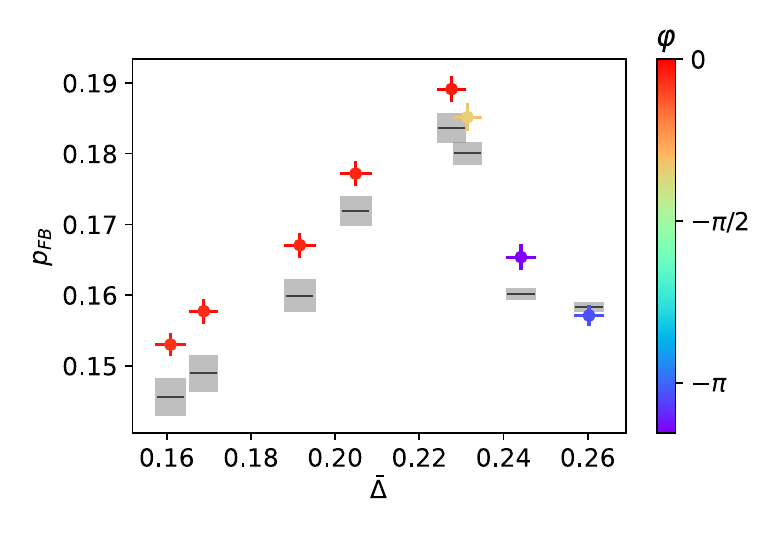}
    \caption{\textbf{Counter-intuitive behaviour of full bunching probabilities.} Experimentally measured full-bunching probabilities $\Tilde{p}_{\text{FB}}$ at the output of a balanced Fourier interferometer (colored crosses), as a function of the measured average overlap $\bar{\Delta}_{\text{exp}}$, showing that one can have decreasing probability of full bunching as the average overlap increases. The colormap shows the triad phase $\varphi$ of the corresponding preparation. The measured results are compared to a numerical simulation that considers the experimentally measured Gram matrices.}
    \label{fig:pfb_behaviour}
\end{figure}
If one fixes $\delta = \frac{\pi}{6}$, for any choice of the polarization phase $\phi$ one can also tune the time-delay parameter $x$ so that $\sqrt{\Delta_{ij}} \approx \frac{1}{2}$. By keeping fixed the average overlap $\bar{\Delta}_{(\phi,x)}$, as in the previous section, one obtains a full bunching probability monotonically increasing with $\phi$. Again such choice of $\phi$ is in a one-to-one correspondence with the induced triad phase $\varphi$. Interestingly, given a fixed value of $\phi$ and $x$, by increasing $\delta$ up to the point in which the 3-state average overlap $\bar{\Delta}_{(\phi,x)}(\delta)$ attains its minimum, $p_{\text{FB}}(\phi,x,\delta)$ behaves as a monotonically decreasing quantity with $\delta$. Overall, this procedure gives us experimental access to Gram matrices whose $p_{\text{FB}}$ decreases with increasing average overlap $\bar{\Delta}$. In Fig.\ \ref{fig:pfb_behaviour}, we report an experimental measurement of such behaviour, where we show the experimentally measured $\Tilde{p}_{\text{FB}}$ as a function of the measured average overlap $\bar{\Delta}_{\text{exp}}$. For the five left-most points in the Figure, which have been prepared with a fixed phase $\phi = \varphi \approx 0$, we find an intuitive monotonically increasing behaviour of $p_{\text{FB}}(\bar{\Delta})$. However, by changing also the value of $\varphi$ through the polarization phase $\phi$ one can observe a trend where the full-bunching probability decreases for increasing values of $\bar{\Delta}$ (the respective values of all relevant measured quantities and additional information can be found in Supplementary Note 6). Thus, our results show that by careful engineering of indistinguishability scenarios involving both polarization and time degrees of freedom, one can observe a counter-intuitive behaviour of the full bunching probability. 

\section{Discussion}

In this work, we experimentally investigated counter-intuitive behaviour of (full) bunching probabilities in 3-photon interference experiments with a balanced tritter. We observed that increasing distinguishability can lead to a larger probability of bunching events, where two or more photons coalesce in the same output mode. This can be associated to specific symmetries of the tritter which forbid outcomes with two photons in the same output mode if the input photons are fully indistinguishable \cite{tichy2010zero}. We have also experimentally verified the counter-intuitive behavior of full bunching, i.e events with 3 photons in the same output,  a more challenging experiment requiring a careful manipulation of pairwise overlaps between photonic wavefunctions describing polarization and time degrees of freedom. This enabled the demonstration of a scenario where an increase of pairwise indistinguishability together with a precise engineering of a triad phase \cite{menssen2017distinguishability} actually resulted in a decrease of the full bunching probabilities. Our work highlights the complex interplay between bosonic indistinguishability and boson bunching phenomena, the understanding of which is crucial for the development of photonic quantum technologies such as quantum computing and quantum metrology. 

\section*{Acknowledgments}
This work was supported by the FET project PHOQUSING (“PHOtonic Quantum SamplING machine” - Grant Agreement No. 899544) and by ICSC – Centro Nazionale di Ricerca in High Performance Computing, Big Data and Quantum Computing, funded by European Union – NextGenerationEU. LN and EFG acknowledge support from FCT -- Fundação para a Ciência e a Tecnologia (Portugal) via project CEECINST/00062/2018. CF acknowledges support from FCT -- Fundação para a Ciência e a Tecnologia (Portugal) via PhD Grant SFRH/BD/150770/2020. DJB acknowledges support from Brazilian funding agencies FAPERJ and CNPq.

\section*{Methods}

\textbf{\textit{Experimental apparatus.}} As stated in the main text, the experimental apparata can be divided into three sequentially connected stages: generation of single-photon states, their manipulation with a time-to-spatial demultiplexing setup to generate a chosen 3-photon Gram matrix, state evolution and reconstruction of the output probabilities obtained by suitably reconfiguring an integrated photonic interferometer.

{\em Single photon source - }In this work single photons are generated via a Quantum Dot (QD) based source \cite{heindel2023quantum, senellart2017high}. We employ an InGaAs QD \cite{gazzano2013bright,somaschi2016near,ollivier2020reproducibility} operated at a repetition rate of $RR \approx 79$ MHz in the so-called Longitudinal Acoustic (LA) phonon-assisted excitation scheme \cite{thomas2021bright}. In this non-resonant excitation scheme, a pulsed pump laser is shone on the source with a wavelength of $927.2$ nm, slightly blue-detuned from the relevant QD excitonic state - found to be at $927.8$ nm. In such a way, filtering between the emitted single photons and the residual pump laser can be obtained via a set of three band-pass filters set at the single photon wavelength \cite{thomas2021bright}. At the output of the single-photon source, we measured on avalanche photodiode detectors (APDs) a single-photon rate of $\sim 3.5$ MHz, achieving with a single photon purity of $ g^{(2)}(0) \sim 0.017(3)$ evaluated via a standard Hanbury-Brown-Twiss setup. The measured pairwise photon indistinguishability at source level was found to be $V_{\text{HOM}} \sim 0.90(1)$, measured via a Hong-Ou-Mandel experiment implemented in a time-unbalanced Mach-Zehnder interferometer \cite{hong1987measurement} by interfering photons separated by $\Delta t = \frac{1}{RR}$.

{\em Three-photon resource preparation - } Then, the single photon stream is directed to a time-to-spatial demultiplexing setup with a single-mode fiber. In this stage photons are initially steered, by means of an acousto-optical-modulator, into one output channel at a time for a time duration of $T \sim 180$ ns, resulting in a total of three different occupied spatial modes. Thereafter, fiber delays are employed for the synchronization of the single photons on the different output channels of the device, akin to the scheme described in \cite{anton2019interfacing, Pont22, pont2022high, rodari2024semi}. Note that in order to accomplish the task of this experiment, a polarization and a time delay control of each photon is implemented in this stage.

{\em Gram matrix preparation - } Indeed, for any Gram matrix of dimension $n$ it is always possible to find a set of $n$ quantum states $\{\ket{\psi_j}\}$, such that $G_{ij}=\bra{\psi_i}\ket{\psi_j}$. One way to obtain such \emph{quantum realization} of a Gram matrix is via the so-called Cholesky decomposition \cite{horn2012matrix}. Namely, the rank $r$ of a Gram matrix corresponds to the dimension of the vector space spanned by the states $\{\ket{\psi_j}\}$, and there exist a unique decomposition 
\begin{equation}
G=LL^{\dagger}
\end{equation}
where $L$ is a lower-triangular matrix. From the rows of $L$, we can read out the entries of the quantum states realizing the desired Gram matrix. For $n=3$, the fact that $L$ is lower triangular means that these have the form:
\begin{align}
    \ket{\psi_1}&=\ket{1} \nonumber\\
    \ket{\psi_2}&=\alpha_1 \ket{1}+ \alpha_2\ket{2} \nonumber \\
\ket{\psi_3}&=\beta_1 \ket{1}+ \beta_2\ket{2}+\beta_3\ket{3} \nonumber, 
\end{align}
if the matrix has full rank. If it has rank $r=2$, then $\beta_3=0$. As said, a representation of this form (resulting in a lower-triangular $L$) is unique. However, if we apply the same unitary $U$ to all the states, we obtain a different physical realization of the same Gram matrix.

To physically realize the three-photon states necessary to generate any given $n=3$ Gram matrix, we employed polarization and time delay control to achieve tuning of the state $\{\ket{\psi_j}\}$ of each component comprising our three-photon resource. Specifically, polarization control consists of a linear polarizer and a half-wave plate (HWP), together with a liquid crystal (LC) retarder placed in the third output of the DMX. Here, an additional tunable delay line is inserted to reduce, if needed, the degree of distinguishability in the time degree of freedom. In such a way, we can prepare states with the general form of Eq. \eqref{eqn:stateprep}.  Physically, tuning of the polarization parameters  $\{\alpha, \beta, \gamma\}$ is obtained via the HWPs; tuning of $\phi$ is induced by the LC and finally, the parameter $x$ can be modulated via the time-delay-line.

{\em State evolution and measurement - } Then, the three photon resource with each photon prepared in state $\{\ket{\psi_j}\}$ is injected in modes $\{3,4,5\}$ of an eight-mode fully reconfigurable integrated photonic processor (IPP) \cite{rodari2024semi}, obtained via femto-second laser writing \cite{ceccarelli2020low}. The IPP is composed by a mesh of 28 Mach-Zehnder interferometers, arranged in a rectangular shape which has been shown \cite{Clements2016} to be able to implement any unitary transformation between input and output modes. Our IPP is built to operate in a polarization-insensitive fashion \cite{pentangelo2024high} meaning that the unitary transformation implemented by the circuit does not change whatever polarization state is injected at the input. This unique property of femtosecond-laser-written photonic circuits decouples the polarization and the path degrees of freedom of the photons, and thus one can use the former to prepare the desired Gram matrix without affecting the tritter and pseudo-number-resolving transformations implemented on chip. Programmability of the circuit with high-fidelity to any given target unitary $U_8$ is obtained via thermo-optical phase shifters \cite{flamini2015thermally} whose induced optical response has been characterized in order to be able to implement any given transformation.
The experiment involved programming the device in two different blocks, as described in the main text.

We characterized the effective unitary $\Tilde{U}_3$ being programmed into the device with standard methods \cite{laing2012super} relying on the measurement of the pairwise output visibility when injecting a two photon state in the input of the interferometer and adapted to our setup as already described in \cite{rodari2024semi}. We found the moduli of the effective unitary to be:
\begin{equation}
    |\tilde{U}_3| =
    \begin{pmatrix}
0.5998(9) & 0.5563(9) & 0.575(1) \\
0.5471(8) & 0.5906(9) & 0.593(1)\\
0.584(1) & 0.585(1) & 0.563(1)
\end{pmatrix}
\end{equation}
with phases
\begin{equation}
    \arg(\tilde{U}_3) = 
    \begin{pmatrix}
0 &0 & 0 \\
0 & 2.135(4) & -2.106(4) \\
0 & -2.081(4) &  2.158(4)
\end{pmatrix},
\end{equation}
achieving a fidelity with the ideal balanced three-mode Fourier interferometer:
\begin{equation}
U_3 = \frac{1}{\sqrt{3}}
    \begin{pmatrix}
        1& 1& 1  \\
        1& e^{i 2\pi/3}& e^{i 4\pi/3} \\
        1& e^{i 4\pi/3}& e^{i 8\pi/3} 
    \end{pmatrix}
\end{equation}
of $\mathcal{F} = \vert \Tr\{U_3^\dag \Tilde{U}_3\} \vert/3 = 0.99922(4)$.


%

\end{document}


\title{Supplementary Information: Experimental observation of counter-intuitive features of photonic bunching}

\author{Giovanni Rodari}
\address{Dipartimento di Fisica, Sapienza Universit\`{a} di Roma, Piazzale Aldo Moro 5, I-00185 Roma, Italy}

\author{Carlos Fernandes}
\address{International Iberian Nanotechnology Laboratory (INL) Av. Mestre José Veiga s/n, 4715-330 Braga, Portugal}

\author{Eugenio Caruccio}
\address{Dipartimento di Fisica, Sapienza Universit\`{a} di Roma, Piazzale Aldo Moro 5, I-00185 Roma, Italy}

\author{Alessia Suprano}
\address{Dipartimento di Fisica, Sapienza Universit\`{a} di Roma, Piazzale Aldo Moro 5, I-00185 Roma, Italy}

\author{Francesco Hoch}
\address{Dipartimento di Fisica, Sapienza Universit\`{a} di Roma, Piazzale Aldo Moro 5, I-00185 Roma, Italy}

\author{Taira Giordani}
\address{Dipartimento di Fisica, Sapienza Universit\`{a} di Roma, Piazzale Aldo Moro 5, I-00185 Roma, Italy}

\author{Gonzalo Carvacho}
\address{Dipartimento di Fisica, Sapienza Universit\`{a} di Roma, Piazzale Aldo Moro 5, I-00185 Roma, Italy}

\author{Riccardo Albiero}
\address{Istituto di Fotonica e Nanotecnologie, Consiglio Nazionale delle Ricerche (IFN-CNR), Piazza Leonardo da Vinci 32, I-20133 Milano, Italy}

\author{Niki Di Giano}
\address{Istituto di Fotonica e Nanotecnologie, Consiglio Nazionale delle Ricerche (IFN-CNR), Piazza Leonardo da Vinci 32, I-20133 Milano, Italy}
\address{Dipartimento di Fisica, Politecnico di Milano, Piazza Leonardo da Vinci 32, 20133 Milano, Italy}

\author{Giacomo Corrielli}
\address{Istituto di Fotonica e Nanotecnologie, Consiglio Nazionale delle Ricerche (IFN-CNR), Piazza Leonardo da Vinci 32, I-20133 Milano, Italy}

\author{Francesco Ceccarelli}
\address{Istituto di Fotonica e Nanotecnologie, Consiglio Nazionale delle Ricerche (IFN-CNR), Piazza Leonardo da Vinci 32, I-20133 Milano, Italy}

\author{Roberto Osellame}
\address{Istituto di Fotonica e Nanotecnologie, Consiglio Nazionale delle Ricerche (IFN-CNR), Piazza Leonardo da Vinci 32, I-20133 Milano, Italy}

\author{Daniel J. Brod}
\address{Instituto de F\'{i}sica, Universidade Federal Fluminense, Niter\'{o}i -- RJ, Brazil}

\author{Leonardo Novo}
\address{International Iberian Nanotechnology Laboratory (INL) Av. Mestre José Veiga s/n, 4715-330 Braga, Portugal}

\author{Nicol\`o Spagnolo}
\email{nicolo.spagnolo@uniroma1.it}
\address{Dipartimento di Fisica, Sapienza Universit\`{a} di Roma, Piazzale Aldo Moro 5, I-00185 Roma, Italy}

\author{Ernesto F. Galv\~{a}o}
\email{ernesto.galvao@inl.int}
\address{International Iberian Nanotechnology Laboratory (INL) Av. Mestre José Veiga s/n, 4715-330 Braga, Portugal}
\address{Instituto de F\'{i}sica, Universidade Federal Fluminense, Niter\'{o}i -- RJ, Brazil}

\author{Fabio Sciarrino}
\address{Dipartimento di Fisica, Sapienza Universit\`{a} di Roma, Piazzale Aldo Moro 5, I-00185 Roma, Italy}

\maketitle

\section{General Remarks about the Gram Matrices}\label{sec:remarks_Gram}

In what follows, we provide some general remarks about the space of reachable Gram matrices in a 3-photon setup. These are some general properties about Gram matrices~\cite{horn2012matrix}:

\begin{itemize}
    \item A 3x3 Gram matrix $G$ is described by the inner products amongst a set of three vectors, i.e. quantum states $\ket{\psi_{1,2,3}}$. A Gram matrix is then invariant upon applying a unitary transformation to its generating vectors.
    \item A Gram matrix, generated by vector defined on a complex field, is a Hermitian and positive semi-definite (PSD) matrix.
\end{itemize}

The third condition becomes interesting when trying to find a physical realization of a generic Gram matrix $G$. By being Hermitian, a Gram matrix must have real eigenvalues $\lambda_i \in \mathcal{R}$; since it is also a PSD matrix, such eigenvalues must be also positive $\lambda_i \geq 0$. Since the determinant of a matrix can be written as the product of its eigenvalues, if we write:
\begin{equation}
    G = \begin{pmatrix}
1 & x & y \\
x & 1 & ze^{i\varphi}\\
y & ze^{-i\varphi} & 1
\end{pmatrix};
\end{equation}
a necessary condition for its physical realizability is that:
\begin{equation}
    \begin{split}
        \det(G) = 1 + 2xyz\cos(\varphi) - x^2 - y^2 - z^2 \geq 0
    \end{split}
\end{equation}
which in turn gives an algebraic inequality for the feasible values of the triad phase $\varphi$:
\begin{equation}
    \cos(\varphi) \geq \frac{x^2 + y^2 + z^2 - 1}{2xyz}
\end{equation}
The equality in the above condition is what traces the droplet-shaped boundary in the second panel of Fig. \ref{fig:pfb_behaviour_si}, as also discussed in more detail in \cite{2024unitaryinvariant}. Restricting ourselves to a set of ``balanced'' Gram matrices, i.e., such that $x=y=z$, this means that:
\begin{equation}
    \cos(\varphi) \geq \frac{3x^2 - 1}{2x^3}
\end{equation}
which can be always fulfilled iff $0 < x \leq 0.5$.

\section{Bunching probabilities in random and balanced interferometers}

It is natural to ask how easy it is to find interferometers where bunching probability is not maximized for fully indistinguishable photons. In this Section we consider this question for larger system sizes than the one implemented experimentally, where $n=3$. We restrict ourselves to the case where the input to the interferometer consists of a single photon per mode (and thus the numbers of photons and of modes are equal), and leave it as an interesting open question a full investigation when some modes can be initialized with no photons. We also focus on the specific case of comparing bunching probabilities between fully indistinguishable and fully distinguishable bosons. 

We generated $100001$ Haar-random unitary matrices of size $n\times n$ for each $n$ from $3$ to $9$ and
counted how many demonstrate counter-intuitive behaviour, i.e. when the probability of bunching is higher for distinguishable particles than for indistinguishable ones. 

The number of matrices of each dimension resulting in counter-intuitive behaviour are reported in Table \ref{tab:dimension}. For matrices of dimension $6$ or greater no matrices displaying counter intuitive behaviour were generated, suggesting such matrices become vanishingly rare even for systems of modest size.
\begin{table}[ht!]
\begin{tabular}{|c|c|c|c|c|}
\hline 
dimension & 3 & 4 & 5 & $\geq6$\tabularnewline
\hline 
$\#$ of matrices & 1890 & 217 & 17 & 0\tabularnewline
\hline 
\end{tabular}
\caption{\textbf{Number of matrices resulting in counter-intuitive behaviour.} Results of a numerical simulation showing the number of matrices showing counter-intuitive features, from a sample of $100001$ Haar-random unitaries, as a function of the number of modes $m$ (dimension).}
\label{tab:dimension}
\end{table}

In spite of the rarity of counter-intuitive behaviour outside of small systems, there are certain families of matrices which display counter-intuitive behaviour even for larger dimension. An example of such matrices are Fourier matrices $F_n$ for certain odd dimensions $n$ (for even dimensions the permanent vanishes and the bunching probability for indistinguishable photons is 1). The values of the bunching probabilities for distinguishable and indistinguishable photons are given by $p^{\text{(dist.)}}_{\text{B}}= 1- n!/n^n$ and $p^{\text{(ind.)}}_{\text{B}}= 1- |\text{Per}(F_{n})|^2$, respectively. As these values tend both to 1 asymptotically, in Table~\ref{table:Fourier_distvsindist} we present the values of the \textit{antibunching probabilities}, that is the probability of event (11...1) comparing the cases of distinguishable vs indistinguishable photons. We can find counter-intuitive behavior for dimensions 3, 7, 11, 13, 17 and 21.

\begin{table}[h!]
\begin{tabular}{|c|c|c|c|c|}
\hline 
n & Distinguishable & Indistinguishable & Gap & Indist./Dist. Ratio\tabularnewline
\hline 
3 & 0.2222 & 0.333 & -0.1111 & 1.5\tabularnewline
\hline 
5 & 3.80e-2 & 8.000e-3 & 3.040e-2 & 0.2083\tabularnewline
\hline 
7 & 6.120e-3 & 1.339e-2 & -7.267e-3 & 2.188\tabularnewline
\hline 
9 & 9.367e-4 & 1.694e-5 & 9.197e-4 & 0.01808\tabularnewline
\hline 
11 & 1.399e-4 & 1.604e-4 & -2.050e-5 & 1.147\tabularnewline
\hline 
13 & 2.056e-5 & 1.020e-4 & -8.142e-5 & 4.960\tabularnewline
\hline 
15 & 2.986e-6 & 2.107e-9 & 2.984e-6 & 7.057e-4\tabularnewline
\hline 
17 & 4.300e-7 & 7.689e-7 & -3.389e-7 & 1.788\tabularnewline
\hline 
19 & 6.149e-8 & 1.032e-8 & 5.116e-8 & 0.1679\tabularnewline
\hline 
21 & 8.745e-9 & 3.54e-9 & 5.204e-9 & 0.4049\tabularnewline
\hline 
\end{tabular}
\caption{\textbf{Probability of antibunching events, i.e. outcome (11\ldots 1),  in Fourier interferometers with one photon per input mode, in the case of distinguishable and indistinguishable photons.} The difference between the values and their ratio is also presented. A negative gap value (or a ratio larger than 1), means that $p_{\text{dist.}}(11\ldots 1)<p_{\text{indist.}}(11\ldots 1)$ showing examples of counter-intuitive behavior where distinguishable particles lead to a higher bunching probability  than indistinguishable bosons.}
\label{table:Fourier_distvsindist}
\end{table}
Exploiting existing work on permanents of real-valued Hadamard matrices, it is also possible to compute bunching probabilities for Hadamard interferometers, a balanced $n$-mode interferometer with entries $\pm 1/\sqrt{n}$. These matrices only exist if $n\leq 2$ or $n$ divides $4$. For a given dimension there may exist several different equivalence classes of Hadamard matrices with different values of the permanent. These values have been computed explicitly in \cite{wanless2005permanents} for all real-valued Hadamard matrices up to dimension 28. By taking the equivalence class with the largest value for the permanent in a given dimension, it is possible to find examples where  $p^{\text{(dist.)}}_{\text{B}} >p^{\text{(ind.)}}_{\text{B}}$ for dimensions 4, 8, 12, 16, 20, 24 and 28 (see Table~\ref{table:bunching_Had}).   
\begin{table}
\begin{tabular}{|c|c|c|c|c|}
\hline 
n & Distinguishable & Indistinguishable & Gap & Indist./Dist. Ratio\tabularnewline
\hline 
2 & 0.5000 & 0 & 0.500 & 0\tabularnewline
\hline 
4 & 9.375e-2 & 0.2500 & -0.1563 & 2.667\tabularnewline
\hline 
8 & 2.403e-3 & 8.789e-3 & -6.386e-3 & 3.656\tabularnewline
\hline 
12 & 5.372e-5 & 2.3814e-4 & -1.844e-4 & 4.433\tabularnewline
\hline 
16 & 1.134e-6 & 1.393e-4 & -1.382e-4 & 122.8\tabularnewline
\hline 
20 & 2.320e-8 & 4.591e-10 & 2.274e-8 & 1.979e-2\tabularnewline
\hline 
24 & 4.652e-10 & 1.461e-8 & -1.412e-8 & 31.40\tabularnewline
\hline
28 & 1.34877e-11 & 9.19848e-12 & -4.28922e-12 & 1.4663\tabularnewline
\hline 
\end{tabular}
\caption{\textbf{Probability of antibunching events, i.e. outcome (11\ldots 1),  in selected examples of real-valued Hadamard interferometers with one photon per input mode, for the cases of perfectly distinguishable and perfectly indistinguishable photons.} For each dimension we choose the real-valued Hadamard interferometer with the largest permanent using the values listed in \cite{wanless2005permanents}. The difference between the probabilities and their ratio is also presented. A negative gap value (or a ratio larger than 1), means that $p_{\text{dist.}}(11\ldots 1)<p_{\text{indist.}}(11\ldots 1)$ showing examples of counter-intuitive behavior where distinguishable particles lead to a higher bunching probability  than indistinguishable bosons.}
\label{table:bunching_Had}
\end{table}

Finally, we point out that though our numerical investigation only revealed this counter-intuitive behaviour up to small matrix dimensions, it can be shown to happen for arbitrarily large system sizes. A trivial construction that achieves this is a $3n$-mode interferometer consisting of $n$ parallel tritters, though the corresponding gap is exponentially supressed. If for each tritter the probability of observing outcome $(111)$ is $0.2222$ (resp.\ $0.3333$) in the distinguishable (resp.\ indistinguishable) case, then the probability of this happening coincidentally for all $n$ tritters is $0.2222^n$ (resp.\ $0.3333^n$), with a gap of $0.2222^n-0.3333^n$.

\section{Maximum and minimum of the bunching probability}
In this section, for ease of notation we will denote the Bargmann invariant as $\Delta_{123}= r e^{i \varphi}$.  The theorem stating that the arithmetic mean must be greater or equal to the geometric
mean gives us the relation
\begin{equation}
r^{\frac{2}{3}}\leq\bar{\Delta}. 
\end{equation}
In turn,  the requirement that the Gram matrix be positive definite
implies its determinant must be positive and therefore
\begin{equation}
\bar{\Delta}\leq\frac{1+2 r\cos\varphi}{3}.
\end{equation}
The last two results provide strict upper and lower bounds for the
average overlap $\bar{\Delta}$ for a given absolute value of the Bargmann invariant $r$:
\begin{equation}
r^{\frac{2}{3}}\leq\bar{\Delta}\leq\frac{1+2r\cos\phi}{3}.
\end{equation}
Replacing in the expression for the bunching probability we have

\begin{equation}
\frac{7-4r\cos\phi+3r^{\frac{2}{3}}}{9}\leq p_{\text{B}}\leq\frac{8-2r\cos\phi}{9}
\end{equation}
We can find the greatest bunching probability by maximizing the upper
bound. This can be done by minimizing the real part of the Bargmann invariant, which is achieved for $r=1/8$ and $\phi=\pi$ \cite{2024unitaryinvariant}, corresponding to a bunching probability $p_{\text{B}}= 11/12$.
The minimum of the bunching probability is analogously achieved by 
minimizing
the lower bound. For any value of $r$, we can decrease the lower bound by setting $\varphi=0$. Hence the minimum is obtained as 
\begin{equation}
P_{\min}=\min_{0\leq r\leq 1}\left(\frac{7-4r+3r^{\frac{2}{3}}}
{9}\right).
\end{equation}
This is minimized for $r=1$, corresponding to fully indistinguishable photons, and a minimum bunching probability of $p_{\text{B}}=2/3$.

\section{Experimental setup}

As stated in the main text, in order to engineer a Gram matrix different from the one where all entries are equal to 1, i.e. the perfectly indistinguishable case, one must be able to control the input photons' spectral functions $\ket{\psi_i}$, which are then injected into an optical interferometer in which the photons interfere. Then, one must also be able to reconstruct the full probability distribution - in the photon number space - at the output of the interferometer.
The experimental setup can be divided into three sequentially connected stages, related to single-photon generation via a Quantum Dot-based source; multi-photon state preparation with a bulk time-to-spatial demultiplexing setup; and state evolution with pseudo-photon-number resolved detection implemented via an integrated reconfigurable interferometric mesh.

A stream of single photons is generated via an InGaAs Quantum Dot (QD) \cite{heindel2023quantum, senellart2017high,gazzano2013bright,somaschi2016near,ollivier2020reproducibility} operated at a repetition rate of $\approx 79$ MHz. More specifically, we employ the so-called Longitudinal Acoustic (LA) phonon-assisted excitation scheme \cite{thomas2021bright}: a pulsed pump laser is shone on the source with a wavelength of $927.2$ nm, slightly blue-detuned from the relevant QD excitonic state - found to be at $927.8$ nm. In such a way, filtering between the emitted single photons and the residual pump laser can be obtained via a set of three band-pass filters set at the single photon wavelength. At the output of the source, we measured on avalanche photodiode detectors a single-photon rate of $\approx 3.5$ MHz, achieving with a single photon purity of $ g^{(2)}(0) \approx 0.02$ evaluated via a standard Hanbury-Brown-Twiss setup together with a pairwise photon indistinguishability of $V_{\text{HOM}} \approx 0.90$ measured via a Hong-Ou-Mandel experiment implemented in a time-unbalanced Mach-Zehnder interferometer \cite{hong1987measurement}.

Then, the single photons are directed to a time-to-spatial demultiplexing setup \cite{anton2019interfacing, Pont22, pont2022high, rodari2024semi} via a single-mode fiber. In this stage, the photons are initially steered, by means of an acousto-optical-modulator, into one output channel at a time for a time duration of $T \sim 180$ ns, resulting in a total of three different occupied spatial modes. Thereafter, fiber delays are employed for the synchronization of the single photons on the different output channels of the device. Moreover, a polarization and a time delay control are implemented in this stage in order to accomplish the task of this experiment. Specifically, the polarization control consists of a polarizer and a waveplate $\lambda/2$ (respectively, purple and green rectangles in Fig. 2 of the main text) on each channel, a liquid crystal (yellow rectangle in Fig. 2 of the main text) on the same channel of the subsequently time delay line.

At this point, the three trains of single photons, each associated with a certain spectral function, are injected into an eight-mode fully reconfigurable integrated photonic processor (IPP) with rectangular shape \cite{Clements2016}. The circuit is fabricated by means of the femtosecond laser writing (FLW) technique \cite{flamini2015thermally} and it is composed of a rectangular mesh of 28 Mach-Zehnder interferometer. Each unit cell consists of two directional couplers, acting as 50:50 beam splitters, and two phase shifters. The phase shifter can be thermally controlled \cite{flamini2015thermally}, allowing the possibility to reconfigure fully the unit cell and, therefore, the behaviour of the entire circuit. The red area in Fig. 2 of the main text is dedicated to the implementation of the three mode balanced Fourier unitary matrix, while the green is involved in Pseudo-Number Resolving detection. Finally, the outgoing photons are detected by avalanche photodiode (APD).

\section{Modeling the experiment}\label{sec: modeling}
Here we discuss a model employed to retrieve the predicted values of the experimentally measured quantities throughout the manuscript, taking into account the main experimental imperfections of the apparatus. The ideal scenario for the experiment corresponds to the implementation of the three-mode Fourier transformation $U_{3}$, with elements $(U_3)_{j,k} = e^{\imath 2 \pi j k/3}/\sqrt{3}$, while measuring the output events for a three-photon input state given by a specific Gram matrix preparation. In the employed apparatus, one needs to consider the presence of some noise sources which modifies the output measurements from the ideal scenario described above (see also \cite{Pont22}).

\textit{Non-ideal implementation of the Fourier interferometer.} In our implementation, the Fourier interferometer is realized by programming the 8-mode universal processor to realize such a transformation. This leads to an effective implemented unitary matrix $\tilde{U}_{3}$. The effective implemented unitary $\tilde{U}_{3}$ is taken into account in the experimental model as the actual transfer matrix between input and output modes.

\textit{Multiphoton contributions from the source.} The second source of noise is related to the multiphoton contributions from the quantum dot source. More specifically, a second photon can be present in each time-bin with a small probability $p^{(2)}$, thus leading to the need of appropriate corrections in the output probabilities. For quantum dot sources \cite{ollivier2021hong}, this noise photon is found to be distinguishable from the main photons. The ratio between the probability of emitting a single photon ($p^{(1)}$) and the one of finding also a noise photon ($p^{(2)}$) can be retrieved from the second order correlation parameter $g^{(2)}(0)$ as $g^{(2)}(0) = 2p^{(1)}/(p^{(1)}+2p^{(2)})^2$. In the complete model for the experiment, we have then taken into account terms due to multiphoton emission by considering all possible contributions to the output probability distributions. 

\textit{Losses.} In our experiment, losses between the different arms are found to be almost balanced. One can thus consider the approximation of balanced losses for our apparatus, and exploit the result of \cite{oszmaniec2018classical}. Losses in the model can be thus made to commute with the interferometer and the demultiplexing module, and placed as a unique loss parameter at the output of the source. As a relevant note, the three-photon experiment is performed in a post-selected scenario, when three photons are measured at the output. Given that multiphoton contributions are present in the source, the role of losses is not limited to a reduction of the detected signal of a factor $\eta^{n}$, being $\eta$ the overall transmission per photon. Indeed, losses also have the effect of changing the relative weights of the different terms due to the presence of multiphoton contributions. The only unbalanced losses contributions are due to detection efficiency associated to the probabilistic photon counting apparatus, and this is taken into account by directly correcting the experimental measured probabilities.

\textit{Estimating the system parameters.} Within the experiment, the parameters to evaluate the predicted results from the model are retrieved as follows. The probabilities $p^{(1)}$ and $p^{(2)}$ are retrieved from the source brightness $B \sim p^{(1)} + p^{(2)}$ and the second order correlation parameter $g^{(2)}(0) = 2p^{(1)}/(p^{(1)}+2p^{(2)})^2$, estimated from a Hanbury-Brown-Twiss experiment on the quantum dot source. Typical values for the $g^{(2)}(0)$ parameter are found in the range $g^{(2)}(0) \in (0.015,0.025)$. Furthermore, losses are directed estimated via a loss budget of the apparatus, leading to an overall efficiency of around $0.011$. We have that the fibered brightness of the source is $\eta_{\text{F}} \approx 0.13$; a transmission efficiency of the DMX setup of $\eta_{\text{DMX}} \approx 0.8$; an efficiency of the polarization filtering of $\eta_{\text{pol}} \approx 0.7$; an integrated chip transmission of $\eta_{\text{chip}} \approx 0.5$ and a detection efficiency of $\eta_{\text{det}} \approx 0.35$. With an additional in-fiber loss of $\eta_{\text{I}} \approx 0.85$ due to splicing and mating sleeves, we have overall: $\eta = \eta_{\text{F}}\eta_{\text{DMX}}\eta_{\text{pol}}\eta_{\text{I}}\eta_{\text{chip}}\eta_{\text{det}} \approx 0.011$.

The effective transformation $\tilde{U}_{3}$ implemented via the reconfigurable processor has been reconstructed via tomographic techniques, where the obtained matrix elements are reported in the Methods. The fidelity between the Fourier transformation $U_{3}$ and the implemented one $\tilde{U}_{3}$ is found to be  $F=\vert \Tr\{U_3^\dag \Tilde{U}_3\} \vert/3 = 0.99922 (4)$, thus showing that the realized evolution is very close to the ideal one leading to very minor corrections in the output probabilities.

Finally, one needs to estimate the effective parameters of the Gram matrix, that is, the two-photon overlaps $\Delta_{ij}$ and the complex phase $\varphi$. The two photon overlaps are estimated by measuring the pair-wise Hong-Ou-Mandel visibilities 
between the photons at the output of the demultiplexer. A second approach has been also used, based on a simple postprocessing of the probabilities of the different outcomes at the output of the interferometer, which allows to estimate both real and imaginary parts of the third-order Bargmann invariant $\Delta_{123}$. More specifically, for a given Gram matrix $G$ as parametrized in the main text, let $p(n_1,  n_2, n_3)$ denote the probability of observing outcome $(n_1, n_2, n_3)$ at the output of the Fourier interferometer. Using the explicit expression of the outcome probabilities from \cite{menssen2017distinguishability} we can define the following probabilities: 
\begin{align}
    P_A& = p(1,1,1)+p(3,0,0)+ p(0,3,0)+ p(0,0,3)= \frac{1}{3}+ \frac{2}{3} |\Delta_{123}|\cos{(\varphi)}\\
    P_B & =  p(0,2,1)+ p(2,1,0)+ p(1,0,2)= \frac{1}{3}- \frac{2}{3} |\Delta_{123}|\cos{(\varphi+ \pi/3)}\\
    P_C & = p(1,2,0)+p(0,1,2)+p(2,0,1)=  \frac{1}{3}- \frac{2}{3} |\Delta_{123}|\cos{(\varphi - \pi/3)}. 
\end{align}
The Bargmann invariant can be extracted from these values via the following relation:
\begin{equation}
    \Delta^N_{123}= |\Delta^N_{123}| e^{i \varphi}= P_A+ P_B e^{i 2 \pi /3 }+ P_C e^{i 4 \pi /3}.  
    \label{eq:bgmn_est}
\end{equation}
From here, it follows that the triad phase associated to a given indistinguishability scenario obtained via modulation of both the polarization and the time degree of freedom of the photonic resources, thus generally associated with a complex-valued Gram matrix $S$, can be estimated experimentally as $\varphi = \arg{\Delta^N_{123}}$.

This set of estimated parameters have been then used to perform the predictions from the model, reported in Figs. 3-5 of the main text. In particular, the two independent estimation of the overlaps and third-order Bargmann invariant have been combined simultaneously to define the indistinguishability properties of the input three-photon states.

\section{Experimental data}
The results regarding the counter-intuitive behavior of photonic bunching were presented in the main text. In this section, we present the measured data used in the figures of the main text. In particular, Tabs. \ref{tab:pol} and \ref{tab:tempo} contain the data of bunching and full bunching probabilities, $p_{\text{B}}$ and $p_{\text{FB}}$, as functions of the mean overlap $\bar{\Delta}$, when operating only on the polarization and time degrees of freedom, respectively. Next, the modulation of bunching and full bunching probabilities, $p_b$ and $p_{fb}$, as the triad phase $\varphi$ varies while keeping the average overlap $\bar{\Delta}$ approximately fixed, is depicted by the data presented in Tab. \ref{tab:lc}. Finally, data describing the counter-intuitive behavior of the full bunching probability $p_{fb}$, which initially increases and then decreases, modulated by the triad phase $\varphi$, as the average overlap $\bar{\Delta}$ increases, are presented in Tab. \ref{tab:counter}. The two-photon overlaps $\Delta_{12}$, $\Delta_{23}$ and $\Delta_{31}$, estimated via the measured visibility $V_{ij}$ of independent Hong-Ou-Mandel experiments corrected for the non-zero multi-photon component estimated via $g^{(2)}(0)$ as in \cite{ollivier2021hong}, are tabulated in all the tables in this section. With this independent measurement, one can estimate the modulus of the Bargmann invariant $\vert\Delta_{123}\vert$ corresponding to a given indistinguishability scenario in two different ways. First, $\vert\Delta_{123}^V\vert$ can be derived from the measured two-photon overlaps as $\sqrt{\Delta_{12}\Delta_{23}\Delta_{13}}$; second, $\vert\Delta_{123}^N\vert$ can be inferred from the experimentally measured photon number distributions with the analytical method presented in Supplementary Note \ref{sec: modeling}. By these two independent estimations of the Bargmann invariant moduli we observe that $\vert\Delta_{123}^N\vert < \vert\Delta_{123}^V\vert$. This bias between the two different estimations can be explained by noting that the output photon number distribution is affected by the presence of multi-photon components, which are corrected in the estimation of $\vert\Delta_{123}^V\vert$, and by the imperfect dialling of the unitary interferometer $\tilde{U}_3$, which slightly differs from an ideal balanced tritter matrix assumed in the derivation of Eq. \eqref{eq:bgmn_est}.

\begin{figure}[ht!]
    \centering
    \includegraphics[width=0.5\columnwidth]{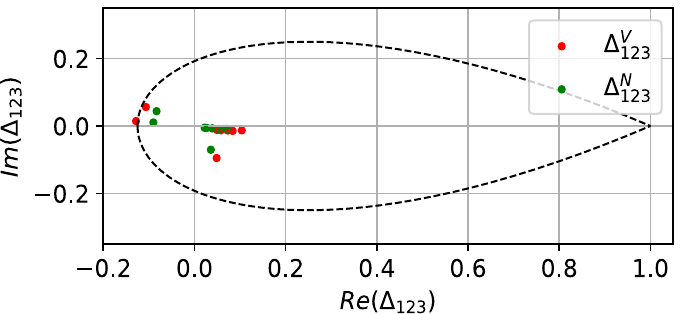}
    \caption{\textbf{Complex plane representation of the measured third-order Bargmann Invariants for the data in supplementary Table \ref{tab:counter}.} Behavior in the complex plane of the experimentally estimated Bargmann invariant parameters $\Delta_{123}$ for each Gram matrix preparation of Table \ref{tab:counter}. Red points correspond to the parameter $\Delta^{N}_{123}$ estimated directly from the probability distributions, as highlighted in Supplementary Note 6. Conversely, green points correspond to the parameter $\Delta^{V}_{123}$ as estimated via an independent measurement of the pairwise photon overlaps $\Delta_{ij}$ together with the triad phase $\varphi$ from the former method. We also report with the black dashed line the border of the physically achievable values of $\Delta_{123}$ \cite{2024unitaryinvariant}, as described in Supplementary Note 1.}
    \label{fig:pfb_behaviour_si}
\end{figure}

Additionally, a complementary picture of the measured points in the main text section \textit{``Experimental observation of counter-intuitive behavior of full bunching''} is shown in Fig. \ref{fig:pfb_behaviour_si}, where the measured Gram matrix preparations are depicted in the complex plane of the estimated third-order Bargmann invariant $\Delta_{123}$ for each experimentally measured configuration.

\begin{table}[ht!]
    \centering
    \begin{tabular}{|ccccc|cc|cc|}
        \hline
        $\Delta_{12}$&$\Delta_{23}$&$\Delta_{31}$&$\bar{\Delta}$&$\vert\Delta_{123}^V\vert$&$\vert\Delta_{123}^N\vert$&$\varphi$&$p_{\text{B}}$&$p_{\text{FB}}$\\
        \hline
        0.269(4)&0.316(4)&0.243 (4)&0.276(3)&0.144(2)&0.092(9)&2.98   (8)&0.893(3)&0.165(6)\\
        0.461(4)&0.340(4)&0.129 (4)&0.310(3)&0.142(3)&0.063(4)&-2.71  (6)&0.883(1)&0.179(3)\\
        0.476(4)&0.422(4)&0.086 (4)&0.328(3)&0.132(4)&0.046(4)&-2.51  (8)&0.880(1)&0.188(3)\\
        0.563(4)&0.570(4)&0.135 (4)&0.423(3)&0.208(4)&0.120(5)&0.126  (3)&0.844(2)&0.257(4)\\
        0.658(4)&0.670(4)&0.220 (5)&0.516(3)&0.312(4)&0.264(5)&0.161  (1)&0.816(2)&0.323(4)\\
        0.595(3)&0.714(3)&0.268 (4)&0.526(2)&0.338(3)&0.276(4)&-0.020 (9)&0.818(1)&0.335(3)\\
        0.769(3)&0.808(3)&0.463 (3)&0.680(3)&0.536(3)&0.476(4)&0.0607 (4)&0.778(2)&0.428(3)\\
        0.853(4)&0.837(4)&0.680 (4)&0.790(3)&0.696(4)&0.642(4)&0.0048 (3)&0.740(2)&0.501(3)\\
        0.897(3)&0.910(3)&0.872 (3)&0.893(2)&0.844(3)&0.785(3)&-0.0084(2)&0.713(2)&0.570(3)\\
        \hline
    \end{tabular}
    \caption{\textbf{Measurements of bunching and full bunching probabilities, $p_{\text{B}}$ and $p_{\text{FB}}$, when only the polarization degree of freedom changes.} The triad phase $\varphi$, the three two-photon overlaps $\Delta_{12}$, $\Delta_{23}$ and $\Delta_{31}$, the average overlap $\bar{\Delta}$, the bunching and full bunching probability, $p_{\text{B}}$ and $p_{\text{FB}}$, the Bargman invariant modulus $\vert\Delta_{123}\vert$, derived with two different methods, are presented here. These measurements are taken for different polarization states that belong to the plane that contains $|H\rangle$, $|V\rangle$, $|D\rangle$ and $|A\rangle$.}
    \label{tab:pol}
\end{table}

\begin{table}[ht!]
    \centering
    \begin{tabular}{|ccccc|cc|cc|}
        \hline
        $\Delta_{12}$&$\Delta_{23}$&$\Delta_{31}$&$\bar{\Delta}$&$\vert\Delta_{123}^V\vert$&$\vert\Delta_{123}^N\vert$&$\varphi$&$p_{\text{B}}$&$p_{\text{FB}}$\\
        \hline
            0.422(4)&0.902(3)&0.405(4)&0.576(3)&0.393(3)&0.355(8)&-0.01(1) &0.796(4)&0.366(7)\\
            0.447(5)&0.861(4)&0.434(5)&0.581(4)&0.409(4)&0.342(9)&-0.03(2) &0.800(4)&0.362(8)\\
            0.495(4)&0.900(3)&0.484(4)&0.626(3)&0.464(4)&0.397(9)&-0.02(1) &0.796(4)&0.394(7)\\
            0.574(4)&0.874(3)&0.539(4)&0.662(3)&0.520(4)&0.454(8)&0.00(1)  &0.780(4)&0.416(7) \\
            0.644(4)&0.882(3)&0.596(4)&0.707(3)&0.582(4)&0.517(8)&0.005(9) &0.767(4)&0.445(7)\\
            0.629(4)&0.894(3)&0.608(4)&0.710(3)&0.584(4)&0.519(8)&0.00 (1) &0.768(4)&0.448(7) \\
            0.715(4)&0.892(4)&0.666(4)&0.758(3)&0.652(4)&0.589(8)&0.003 (8)&0.746(4)&0.472(7)  \\
            0.753(4)&0.878(3)&0.740(4)&0.790(3)&0.699(4)&0.650(7)&-0.006(7)&0.746(5)&0.512(7)\\
            0.808(3)&0.892(3)&0.747(3)&0.815(3)&0.733(4)&0.674(6)&0.012(5) &0.736(4)&0.519(6)\\
            0.886(3)&0.887(3)&0.847(3)&0.873(3)&0.816(4)&0.750(6)&0.0006(4)&0.723(4)&0.557(6)\\
        \hline
    \end{tabular}
    \caption{\textbf{Measurements of bunching and full bunching probabilities, $p_{\text{B}}$ and $p_{\text{FB}}$, when only the time degree of freedom changes.} The triad phase $\varphi$, the three two-photon overlaps $\Delta_{12}$, $\Delta_{23}$ and $\Delta_{31}$, the average overlap $\bar{\Delta}$, the bunching and full bunching probability, $p_{\text{B}}$ and $p_{\text{FB}}$, the Bargman invariant modulus $\vert\Delta_{123}\vert$, derived with two different methods, are presented here. This measurements are taken for different time delay between one channel of the demultiplexer (DMX) and the other two.}
    \label{tab:tempo}
\end{table}

\begin{table}[ht!]
    \centering
    \begin{tabular}{|ccccc|cc|cc|}
        \hline
        $\Delta_{12}$&$\Delta_{23}$&$\Delta_{31}$&$\bar{\Delta}$&$\vert\Delta_{123}^V\vert$&$\vert\Delta_{123}^N\vert$&$\varphi$&$p_{\text{B}}$&$p_{\text{FB}}$\\
        \hline
            0.264(6)&0.277(7)&0.315(6)&0.285(4)&0.152(4)&0.104(9)&-3.13(7)&0.889(3)&0.154(6)\\
            0.268(7)&0.299(7)&0.287(7)&0.285(4)&0.152(4)&0.099(9)&-2.27(9)&0.883(3)&0.174(7)\\
            0.317(6)&0.287(7)&0.299(7)&0.301(4)&0.165(4)&0.119(7)&-1.41(8)&0.845(3)&0.191(7)\\
            0.315(6)&0.277(7)&0.314(6)&0.302(4)&0.165(4)&0.122(8)&-0.85(8)&0.831(4)&0.218(7)\\
            0.286(6)&0.266(7)&0.240(7)&0.264(4)&0.135(4)&0.091(9)&0.015(8)&0.812(4)&0.206(7)\\
            0.330(7)&0.283(7)&0.319(7)&0.311(4)&0.172(4)&0.109(9)&0.574(8)&0.819(4)&0.213(7)\\
            0.302(7)&0.284(7)&0.367(7)&0.317(4)&0.177(4)&0.118(7)&1.25 (8)&0.839(4)&0.198(7)\\
            0.290(7)&0.283(6)&0.276(7)&0.283(4)&0.151(4)&0.096(9)&2.17 (9)&0.869(3)&0.166(7)\\
            0.291(7)&0.283(7)&0.303(6)&0.292(4)&0.158(4)&0.115(9)&2.73 (7)&0.894(3)&0.157(6)\\
        \hline
    \end{tabular}
    \caption{\textbf{Measurements of bunching  and full bunching probabilities, $p_{\text{B}}$ and $p_{\text{FB}}$, when the triad phase $\varphi$ changes by means of a liquid crystal.} The triad phase $\varphi$, the three two-photon overlaps $\Delta_{12}$, $\Delta_{23}$ and $\Delta_{31}$, the average overlap $\bar{\Delta}$, the bunching and full bunching probability, $p_{\text{B}}$ and $p_{\text{FB}}$, the Bargmann invariant modulus $\vert\Delta_{123}\vert$, derived with two different methods,  are presented here. These measurements are taken for different triad phases $\varphi$ exploiting the liquid crystal.}
    \label{tab:lc}
\end{table}

\begin{table}[ht!]
    \centering
    \begin{tabular}{|ccccc|cc|c|}
        \hline
        $\Delta_{12}$&$\Delta_{23}$&$\Delta_{31}$&$\bar{\Delta}$&$\vert\Delta_{123}^V\vert$&$\vert\Delta_{123}^N\vert$&$\varphi$&$p_{\text{FB}}$\\
        \hline
            0.254(6)&0.061(6)&0.168(7)&0.161(4)&0.051(3)&0.023(2)&-0.23(9)&0.153(2)\\
            0.255(6)&0.082(6)&0.169(6)&0.169(3)&0.059(3)&0.028(2)&-0.21(7)&0.158(2)\\
            0.284(7)&0.102(6)&0.189(7)&0.192(4)&0.074(3)&0.040(3)&-0.18(5)&0.167(2)\\
            0.291(6)&0.126(7)&0.198(7)&0.205(4)&0.085(3)&0.055(3)&-0.17(4)&0.177(2)\\
            0.303(6)&0.185(6)&0.196(6)&0.228(4)&0.105(3)&0.072(3)&-0.12(3)&0.189(2)\\
            0.274(6)&0.159(6)&0.261(6)&0.231(3)&0.107(3)&0.079(2)&-1.10(4)&0.185(2)\\
            0.236(7)&0.239(6)&0.258(6)&0.244(4)&0.120(4)&0.094(3)&-3.63(2)&0.165(2)\\
            0.254(6)&0.201(6)&0.326(6)&0.260(3)&0.129(4)&0.091(2)&-3.26(2)&0.157(1)\\
        \hline
    \end{tabular}
    \caption{\textbf{Measurements of counter-intuitive behaviour of the full bunching probability $p_{\text{FB}}$ when $\bar{\Delta}$ increases.} The three two-photon overlaps $\Delta_{12}$, $\Delta_{23}$ and $\Delta_{31}$, the average overlap $\bar{\Delta}$, the modulus of the third-order Bargmann invariant $\vert\Delta_{123}\vert$, derived with two different methods, the full bunching probability $p_{\text{FB}}$ and the triad phase $\varphi$ are presented here.}
    \label{tab:counter}
\end{table}


%